\documentclass [sigconf]{acmart}
\AtBeginDocument{%
  }

\setcopyright{acmlicensed}
\copyrightyear{2018}
\acmYear{2018}
\acmDOI{XXXXXXX.XXXXXXX}
\acmConference[Conference acronym 'XX]{Make sure to enter the correct
  conference title from your rights confirmation email}{June 03--05,
  2018}{Woodstock, NY}
\acmISBN{978-1-4503-XXXX-X/2018/06}

\usepackage{appendix}
\usepackage{tcolorbox}
\usepackage{multirow} 
\usepackage{balance}
\usepackage{algorithm}
\usepackage{diagbox} % 导入diagbox包
\usepackage{array}   % 导入array包以便更好地控制表格
\usepackage{subcaption}
\usepackage{algorithmic}
\usepackage{arydshln}
\usepackage{multirow}
\usepackage{tcolorbox}
\usepackage{xcolor}
\usepackage{multirow}
\usepackage{graphicx}
\usepackage{enumitem}
\usepackage{diagbox}
\usepackage{graphicx} 
\begin{document}

\title{TrackRec: Iterative Alternating Feedback with Chain-of-Thought via Preference Alignment for  Recommendation}


\author{Yu Xia}
\authornote{Both authors contributed equally to this research.}
\orcid{0009-0002-4128-6968}
\affiliation{%
  % \institution{Institute of Software, Chinese Academy of Sciences }
  \institution{University of Chinese Academy of Sciences}
  \city{Beijing}
  \country{China}
}
\email{xiayu24@mails.ucas.ac.cn}

\author{Rui Zhong}
\authornotemark[1]
\affiliation{%
  \institution{Kuaishou Technology}
  \city{Beijing}
  \country{China}
}
\email{zhongrui@kuaishou.com}

\author{Zeyu Song}
\authornotemark[1]
\affiliation{%
  \institution{Kuaishou Technology}
   \city{Beijing}
  \country{China}
}
\email{songzeyu@kuaishou.com}

\author{Wei Yang}
\affiliation{%
 \institution{Kuaishou Technology}
  \city{Beijing}
  \country{China}
}
\email{yangwei08@kuaishou.com}

\author{Junchen Wan}
\affiliation{%
 \institution{Kuaishou Technology}
  \city{Beijing}
  \country{China}
}
\email{wanjunchen@kuaishou.com}

\author{Qingpeng Cai}
\affiliation{%
 \institution{Kuaishou Technology}
  \city{Beijing}
  \country{China}
}
\email{caiqingpeng@kuaishou.com}

\author{Chi Lu}
\affiliation{%
  \institution{Kuaishou Technology}
  \city{Beijing}
  \country{China}
}
\email{luchi@kuaishou.com}

\author{Peng Jiang}
\affiliation{%
  \institution{Kuaishou Technology}
  \city{Beijing}
  \country{China}
}
\email{jiangpeng@kuaishou.com}

\renewcommand{\shortauthors}{Yu Xia et al.}

\begin{abstract}
  The extensive world knowledge and powerful reasoning capabilities of large language models (LLMs) have attracted significant attention in recommendation systems (RS).  Specifically, The chain of thought (CoT) has been shown to improve the performance of LLMs on complex reasoning tasks for RS. However, due to the fact that LLMs often suffer from hallucination issues, there is no guarantee that their reasoning CoT is effective. A key challenge is to further enhance the recommendation capabilities of LLMs through effective CoT reasonings.
% However, there are still challenges in leveraging LLMs to enhance recommendation systems. On one hand, relying solely on zero-shot learning fails to fully utilize the potential of LLMs. On the other hand, LLM reasoning is unconstrained and can sometimes be unreliable. Since chain-of-thought (CoT) has been shown to improve the performance of large models on complex reasoning tasks, we consider introducing CoT into recommendation systems.
Therefore, we propose \textbf{TrackRec}, a framework designed to enhance reasoning capabilities of LLMs for RS. TrackRec specifically focuses on accurately inferring recommendation CoT \textbf{(RecCoT)} for user preference using the knowledge from LLMs. This RecCoT can serve both as an explanation for the LLM's completion of recommendation tasks and as auxiliary features to assist recommendation models in accomplishing recommendation tasks. TrackRec consists of a RecCoT generator $(G)$ and a RecCoT validator $(V)$. Furthermore, we design alternating feedback learning mechanism that $G$ undergoes direct preference optimization via feedback from $V$ to produce increasingly accurate RecCoT aligned with $V$'s standards. Meanwhile, $V$ is fine-tuned using the inference feedback from $G$ to enhance its validation capabilities in alignment with recommendation tasks. Through iterative alternating feedback learning between $G$ and $V$, TrackRec continuously improves the user preference analysis capability of $G$ and the validation capacity of $V$. Extensive experiments demonstrate the effectiveness of our approach, showing that it surpasses state-of-the-art methods. Moreover, TrackRec has been deployed on a lagre advertising platform with hundreds of millions of users, achieving substantial gains.
\end{abstract}




\maketitle

\begin{figure}
\centering
\includegraphics[width=0.4\textwidth]{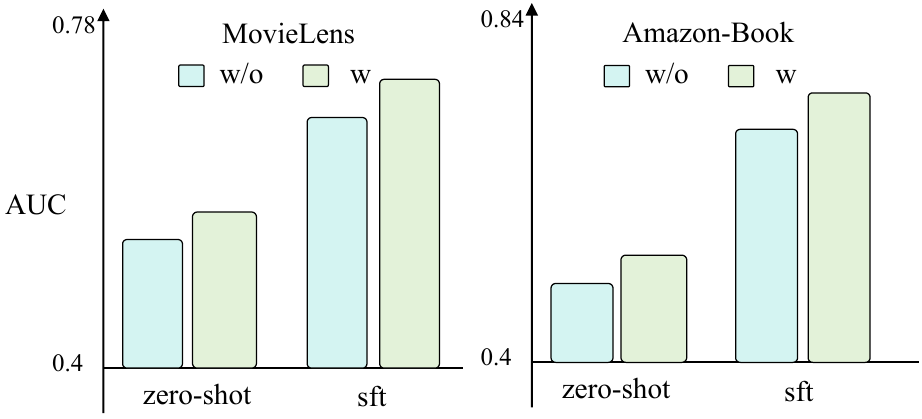}
\caption{The impact of incorporating user preference (RecCoT) on the recommendation performance of LLM when used directly as recommenders. w/o and w refer to without and with RecCoT, respectively.}
\label{fig1}
   \vspace{-3mm}

\end{figure}

\section{Introduction}

Recommendation systems (RS) are ubiquitous in online services, significantly enhancing the user experience in online shopping\cite{w1}, movie discovery \cite{w2}, and music recommendations \cite{w3}, while also greatly boosting the revenue of advertising platforms \cite{w4}. However, traditional RS share a common characteristic—they are trained and deployed on closed and limited datasets, which constrains their performance to some extent.

\begin{figure}[h]
\centering
\includegraphics[width=0.48\textwidth]{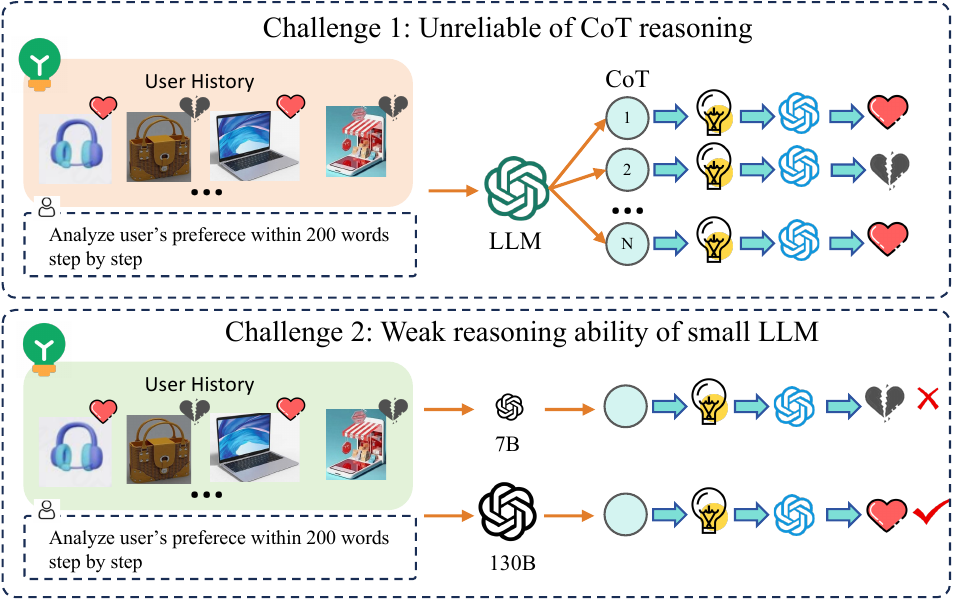}
\caption{ The issues related to introducing chain-of-thought (CoT) into RS, including  unreliable of CoT reasoning and weak reasoning ability of small LLM.}
\label{fig2}
   \vspace{-3mm}
\end{figure}

Large language models (LLMs) have demonstrated strong capabilities in tasks such as text analysis and generating human-like language\cite{w5}, including natural language processing \cite{w6}, data mining \cite{w9, w10, w11}, and information retrieval \cite{w12}. Previous studies have also highlighted the ability of LLMs in knowledge enrichment and combinatorial generalization \cite{w15}. With the right instructions, these models can learn how to solve unseen tasks and activate their knowledge to achieve high performance \cite{w16}. These capabilities of LLMs can effectively bring information gain to RS, alleviating the issue of knowledge gaps. However, due to the complexity of recommendation tasks, it is difficult to achieve strong performance relying solely on zero-shot learning with LLMs. With the introduction of Chain-of-Thought (CoT) \cite{w18}, LLMs are able to deduce more accurate information step by step. This aligns well with our goal of enhancing RS with enriched information from LLMs. As shown in Figure \ref{fig1}, in zero-shot and supervised fine-tuning (sft) setting, we conduct a simple validation of the impact of CoT on RS using typical public datasets, demonstrating the importance of CoT for recommendation tasks. Recently, some works \cite{w24} have attempted to use LLMs as a preprocessing tool for traditional recommendation models, leveraging LLMs' powerful reasoning and analytical capabilities, as well as their extensive knowledge, to infer user summaries that align with RS. While some progress has been made, there are still notable limitations in applying recommendation CoT within RS, as shown in Figure \ref{fig2}: 
\begin{itemize}[leftmargin=*]
\item \textbf{Challenge 1: } Unreliable of CoT reasoning. Due to their pre-training on vast corpora, LLMs produce results with diversity and uncertainty. They may not always derive accurate CoT that satisfy recommendation models, which can affect the final predictions of LLMs. Thus, aligning LLMs' generative space with recommendation systems remains a pressing challenge.
\item  \textbf{Challenge 2: } Weak reasoning ability of small LLM. The knowledge and reasoning capabilities of LLMs are closely related to parameter size. Smaller LLMs cannot fully leverage the advantages of large language models in RS. However, the inference time required by larger LLMs is intolerable in RS. %How to better harness the reasoning capabilities of LLMs is worth further exploration.
\end{itemize}

Considering these challenges, we propose \textbf{TrackRec}, an i\textbf{\underline{t}}e\textbf{\underline{r}}ative \textbf{\underline{a}}lternating feedba\textbf{\underline{ck}} learning framework designed to enhance the reasoning capabilities of LLMs in RS while aligning them with recommendation tasks. We decompose the process of using LLMs for recommendation tasks into two components: a recommendation CoT (RecCoT) generator $(G)$ that generates user preferences and a RecCoT validator $(V)$ that performs the recommendation task based on the user preferences. Both components are derived from a small-scale model, such as Qwen2.5-7B\cite{qwen2}. Furthermore, through direct preference optimization and instruction fine-tuning, we alternately enhance $G$'s user preference reasoning ability and $V$'s validation capability, bridging the gap between LLMs and recommendation tasks. Additionally, we obtain RecCoT generated by a larger LLM through prompt and use it to fine-tuning $G$ for distillation, initializing $G$ to improve the reasoning capabilities of the small-scale model. 

Our contributions are summarized as follows:
\begin{itemize}[leftmargin=*]
\item We validate the importance of RecCoT and propose TrackRec, an iterative alternating feedback learning framework composed of a RecCoT generator $G$ and a RecCoT validator $V$ that continuously enhance RecCoT generation ability and recommendation prediction ability of LLMs. TrackRec maximizes the potential of LLMs in RS, aligning LLMs with downstream recommendation tasks.
\item In process of TrackRec, we design alternating feedback learning mechanism, a preference alignment method to enhance the reasoning capabilities of LLM. Specifically, for $G$, we perform multiple sampling of $G$ to generate various RecCoTs, and use preference optimization based on feedback from $V$ to train $G$. For $V$, we design Rec-tuning. We generate RecCoT through reasoning feedback from the preference-aligned G, and construct instruction data to train V, aligning it with the recommendation task.
% We first obtain RecCoT generated by ChatGPT-4o through prompt and use it to fine-tuning $G$ for distillation. Additionally, we perform multiple samplings on $G$ to generate various RecCoTs, and train $G$ using direct preference optimization based on feedback from $V$. For $V$, we use the inference feedback of $G$ strengthened through preference alignment to generate RecCoTs to construct instruction data for training $V$, thereby aligning with recommendation tasks.

\item  We conduct extensive and comprehensive experiments, verifying the effectiveness and superiority of the proposed framework. Our method achieves state-of-the-art performance on both public and industrial datasets.
\end{itemize}

\begin{figure*}[h]
\centering
\includegraphics[width=0.9\textwidth]{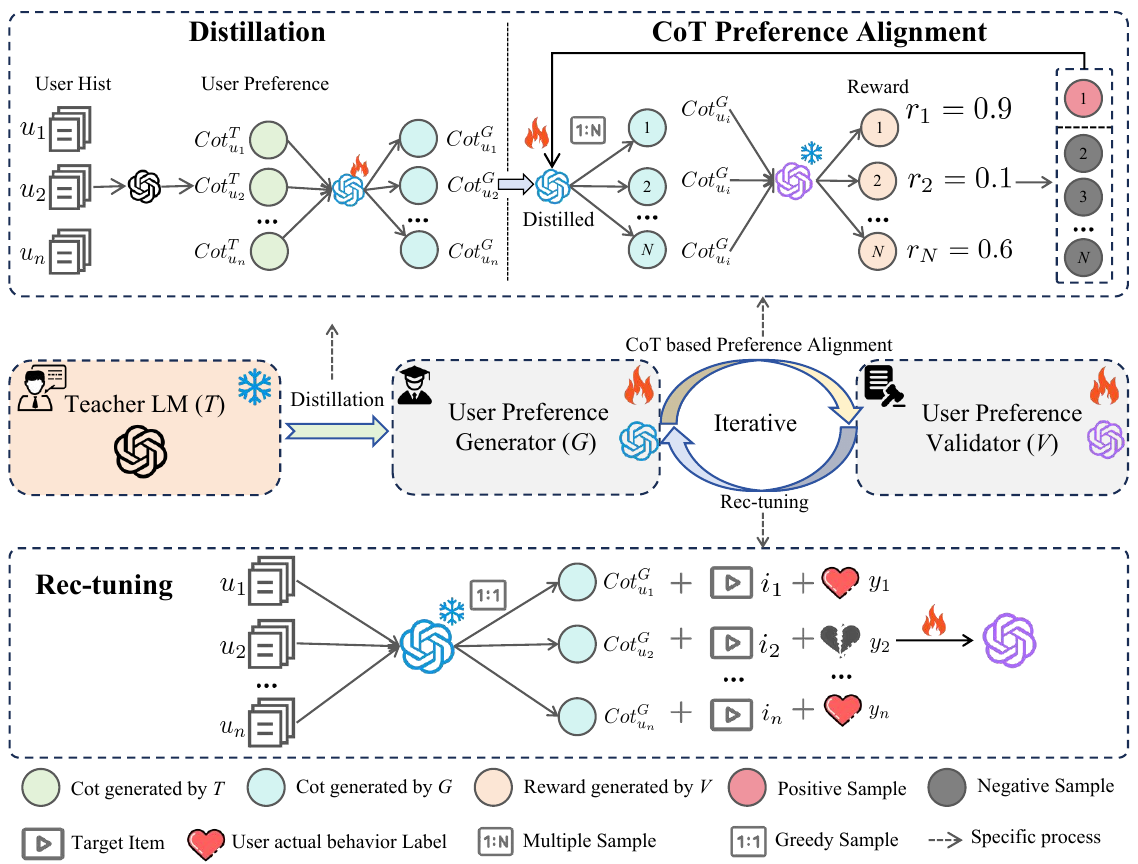}
\caption{The architecture of TrackRec, consists of a recommendation CoT (RecCoT) generator $G$ and a RecCoT validator $V$ along with the four processes: Distillation, CoT Preference Alignment, Rec-tuning and Iterative Alternating Feedback Learning }
\label{fig3}
\end{figure*}

\section{Related Work}
\subsection{LLMs for Recommendation}
The advent of pre-trained language models has brought tremendous success to the field of Natural Language Processing ~\cite{34-bert}, and these models have also demonstrated significant performance within RS ~\cite{32-llmrec,33-llmrec,35-llmrec,67-recapply,70-apply,xia2025hierarchical, zhao2025hierarchical,gu2025r4ec}. 

With the expansion of model parameter scales and corpora, LLMs ~\cite{12_gpt4,36-gpt3.5,61-emergent,63-llmapply} have demonstrated excellent performance across various tasks, particularly in their remarkable reasoning capabilities. Preliminary results have been achieved in studies aiming to apply LLMs within recommendation scenarios ~\cite{37-gpt4rec}. ~\cite{39-chatrec} employs LLMs as an interface for conversational multi-round RS. ~\cite{41-recllm} addresses fairness issues in the application of LLMs for RS. ~\cite{40-zeroshot} further report the zero-shot ranking performance of LLMs with recommender interactions. TALLRec ~\cite{38-tallrec} finetunes LLaMA-7B model with a LoRA architecture on recommendation data. ~\cite{42-baidu} investigates the potential of using LLMs for re-ranking within a recommender context. These preliminary studies have provided insights for applications in industrial settings, yet they have not adequately addressed certain challenges, such as resource issues under large-scale data in industrial environments and inference latency during actual deployment.

\subsection{Reinforcement Learning for LLMs in Recommendation}
Inspired by the exceptional general intelligence of LLMs, researchers have begun to explore the reinforcement learning-based alignment procedure to further strengthen LLMs' aptitude in responding to users' intentions and mitigating formatting errors. 
RLHF ~\cite{rl1} fine-tunes LLMs to generate concise, human-readable user summaries that are optimized for downstream task performance. ~\cite{rl2} propose a reinforcement learning-based alignment procedure to enhance LLMs’ generalization ability.  ~\cite{rl3} enhances the performance of LLM-based recommenders by simultaneously maximizing the probability of positive samples and minimizing the probability of multiple negative samples. 
However, there has been limited exploration of the potential of CoT reasoning in these reinforcement learning-based alignment, which aligns closely with the core objective of recommendations—modeling users' interests in various items. Additionly, the reasoning conducted by LLMs may contradict the actual feedback received in recommendations, and existing methods do not sufficiently address this issue.

\subsection{CoT Reasoning of LLMs}
When addressing complex hierarchical reasoning challenges, the CoT simulates an intuitive cognitive process. These methods ~\cite{49-cot1,50-cot2,51-cot3,66-apply,69-apply} enable models to decompose multi-step problems into intermediate steps. With CoT prompting, LLMs can solve complex reasoning problems that are not solvable with standard prompting methods. Recently, in recommendation scenarios, studies such as InteRecAgent \cite{huang2023recommender} and RecMind \cite{wang2023recmind} have utilized CoT prompting to enable LLMs to function as agents. This approach achieves the management of complex recommendation tasks by breaking them down into sub-tasks and formulating plans that leverage external tools. KAR\cite{14_kar} utilizes prompts to extract world knowledge from LLMs to generate user preferences, thereby enhancing the performance of recommendation models. Current work on CoT reasoning provides valuable insights for us. We need to further optimize the CoT prompts for downstream recommendation tasks.

\section{Methodology}
\subsection{Overview}
In this paper, we propose TrackRec, a recommendation framework designed to enhance the reasoning capabilities of LLMs in RS while aligning them with recommendation tasks. As shown in Figure \ref{fig3}, TrackRec consists of a RecCoT generator $G$ and a RecCoT validator $V$ along with the following four processes: 
\begin{itemize}[leftmargin=*]
    \item \textbf{CoT Preference Alignment}. We perform multiple sampling with $G$ to generate various RecCoTs and align $G$'s preferences based on feedback from $V$.
    \item \textbf{Iterative} alternating feedback learning. We alternately train $G$ and $V$, continuously improving $G$'s reasoning ability in RecCoT and $V$'s recommendation prediction capability. 
    \item \textbf{Rec-tuning}. We generate RecCoTs using the reasoning feedback from the preference-aligned $G$ to construct instruction data for training $V$.
    \item \textbf{Distillation}. We use a larger-scale LLM for distillation to initialize G to enhance the reasoning capabilities of smaller LLMs.
\end{itemize}
% TrackRec have a recommendation chain-of-thought (RecCoT) generator $G$ and a RecCoT validator $V$, which continuously improve their respective abilities through iterative alternating feedback learning. Figure \ref{fig3} illustrates the overall architecture of TrackRec: (1) CoT based Feedback and Preference Alignment, (2) Iterative Alternating Feedback Learning, (3) Rec-tuning and (4) Distillation.

\begin{algorithm}
    \caption{Multiple Sampling Feedback Based on CoT}
    \label{alg1}
    \renewcommand{\algorithmicrequire}{\textbf{Input:}}
\renewcommand{\algorithmicensure}{\textbf{Output:}}
    \begin{algorithmic}
        \REQUIRE{Generator $G$, Validator $V$, User historical interactions set $U = \{u_i\}^{n}_{1}$, Target item set $I = \{i_i\}^{n}_1$, User actual behavior Label Set $Y = \{y_i\}^{n}_{1}$, Number of samples $N$, Sampling Params $t$, $top\_p$\\}
        \FOR{$i \gets 1 $ to $n$}
        % \STATE infer $u_i$ for $N$ times on $G$ and get feedback reward $r$ from $V$
        % $result.setKey((s_i, I_i))$\\
        \FOR{$j \gets 1 $ to $N$}
        \STATE $Cot_{ij} = G(u_i, t, top\_p)$\\
            $a_{ij} = V(u_i,i_i, Cot_{ij})$ \\
            $p^{T}_{ij} = P(a_{ij} = Yes),\ \  p^{F}_{ij} = P(a_{ij} = No)$\\
            $p^{T}_{ij} = \frac{p^{T}_{ij}}{p^{T}_{ij} + p^{F}_{ij}},\ \  p^{F}_{ij} = \frac{p^{F}_{ij}}{p^{T}_{ij} + p^{F}_{ij}}$\\
            $r_{ij} = \begin{cases} 
                        p^{T}_{ij} & \text{if } y_i = Yes \\
                        p^{F}_{ij} & \text{if } y_i = No 
                    \end{cases}$\\    
        % \FOR{$k \gets 1 $ to $H$}
        % \STATE $cluster^{u}_{j,k} = KMeans(e_u)$, $path_u\left[k\right] = j$, $j \in [1,N^k]$\\
        % $cluster^{i}_{j,k} = KMeans(e_i)$, $path_i\left[k\right] = j$, $j \in [1,N^k]$\\
        % \ENDFOR \\
        % \FOR{each $u \in U$}
        % \STATE randomly select $i^{'}$ where $(u,i) \in S $ \AND $(u,i^{'}) \notin S $ \\
        % \STATE \AND $path_{i^{'}}[k] \neq path_{i}[k] (k \in [1,h])$\\
        % \FOR{$k \gets 1 $ to $H$}
        %     add $(u,path_i[k],1)$ \AND $(u,path_{i^{'}}[k],0)$ to $Aug$ \\
        % \ENDFOR \\ 
        \ENDFOR \\
        % $result.setValue((s_i, I_i),\{c_{i*}, r_{i*}\}^{N}_{1})$\\
        % \STATE constructing partial order pairs based on feedback reward $r$
        % \STATE Initialize $R \leftarrow \emptyset$
        % \REPEAT
        % \STATE Sample two elements $(c_{im}, r_{im})$ and $(c_{il}, r_{il})$ from $\{c_i*, r_i*\}$, where $m \neq l$
        % \IF{$r_{im} > r_{il}$}
        %     \STATE Add the partial order pair $(c_{im}, c_{il})$ to $R$
        %     \STATE Return $(c_{im}, r_{im})$ to $\{c_i*, r_i*\}$
        % \ELSE
        %     \STATE Add the partial order pair $(c_{il}, c_{im})$ to $R$
        %     \STATE Return $(c_{il}, r_{il})$ to $\{c_i*, r_i*\}$
        % \ENDIF

        \STATE constructing partial order list based on feedback reward $r$
        % \FOR{$i \gets 1 $ to $n$}
        \STATE \textbf{Initialize:} $C_i \leftarrow \{Cot_{ij}\}, R \leftarrow \{r_{ij}\}$
        \STATE $Cot_{ip} = Cot_{ij} \ where\  r_{ij} = \max\ ( R_i )\ $
        \STATE $C_{i}^{\varepsilon} = C_i - Cot_{ip} $
        % \ENDFOR \\
    % \UNTIL{Only one element left in $\{c_i*, r_i*\}$}
    %     \STATE Initialize $G$ with initial values or structure
    % \FOR{each partial order pair $(c_{p}^*, c_{q}^*) \in R$}
    %     \IF{$G(c_{p}^*) < G(c_{q}^*)$}
    %         \STATE Reinforce $G$ to increase $G(c_{p}^*)$ and/or decrease $G(c_{q}^*)$
    %     \ELSE
    %         \STATE Maintain current structure of $G$ or make minor adjustments
    %     \ENDIF
    % \ENDFOR
    % \STATE Return the reinforced structure $G$
        
        \ENDFOR \\
        \ENSURE{  Positive sample $C_p \leftarrow \{Cot_{ip}\}$\\\ \ \ \ \ \ \ \ \ \ \ \  Negative sample set $C^{\varepsilon} \leftarrow \{C_{i}^{\varepsilon}\}$.} 
        % \vspace{-3mm}
    \end{algorithmic}
\end{algorithm}

% \vspace{-10pt} 

\subsection{CoT Preference Alignment}
To enhance the reasoning capabilities of LLMs in the RS, we design a RecCoT-based approach for $G$, involving multiple sampling and preference alignment based on feedback. As shown in Figure \ref{pprompt}, We construct RecCoT prompt instruction template $prompt^{CoT}$. RecCoT constructs prompts based on items from the user's historical interactions, with the task defined as summarizing user preferences.

\subsubsection{\textbf{CoT based Multiple Sampling Feedback}}

\begin{algorithm}[h]
\caption{Iterative Alterating Feedback Learning for G and V}
\label{alg2}
\begin{algorithmic}
\STATE \textbf{Input:} User historical interactions $U = \{u_i\}_{i=1}^{n}$, Target item $I = \{i_i\}^n_1$, Label Set $Y = \{y_i\}^{n}_{1}$, Maximum iterations $T$, learning rates $\alpha$ and $\beta$, Number of sample $N$.
\STATE \textbf{Initialize:} Generator $G$, Validator $V$.
% \STATE \textbf{Parameters:} , feedback threshold $\tau$, learning rates $\alpha$ and $\beta$.

\FOR{$i \gets 1 $ to $T$}
    \STATE \textbf{Stage 1}: Generator Training (Maximize Feedback Reward) \\
    Do \textbf{Algorithm 1} to get feedback reward $\{C_p\}$ and $C^\varepsilon$
    \FOR{each sample $(u_i, Cot_{p}, C_i^{\varepsilon})$ in $(U, C_{p}, C^\varepsilon)$}
        % \STATE Reforcement learning on $G$
        \STATE  Softmax direct preference optimization on $G$ with objective:
        \[
        \max_G \log (\ \frac{P_G (\boldsymbol{Cot_{p}|u_{i}})^{N-1}}{ \prod_{Cot_d\in{C_i^\varepsilon}}P_G(\boldsymbol{Cot_{d}|u_{i}})}\ )
        \]
        % \STATE  Reinforcement learning loss function can be computed as:
        % \[
        % \begin{aligned}&\mathcal{L}_{\mathrm{RL}}(\pi_\theta;{G})=-\mathbb{E}_{(s_i,(c_w,c_l))\sim\mathcal\{(S,R)\}}\\&\left[\log\sigma\left(\beta\log\frac{\pi_\theta(c_w\mid si)}{{G}(c_w\mid si)}-\beta\log\frac{\pi_\theta(c_{l}\mid s_i)}{{G}(c_{l}\mid s_i)}\right)\right]\end{aligned}
        % \]
        % \STATE \textbf{Step 2:} $V$ evaluates $\hat{y}_G$ and provides feedback score $r = V(\hat{y}_G, y)$.
        % \STATE \textbf{Step 2:} Update $G$ to maximize $V$'s feedback reward by ascending gradient:
        \STATE update $G$'s parameters : 
        \[
        G \leftarrow G + \alpha \nabla \log (\ \frac{P_G (\boldsymbol{Cot_{p}|u_{i}})^{N-1}}{ \prod_{Cot_d\in{C_i^\varepsilon}}P_G(\boldsymbol{Cot_{d}|u_{i}})}\ )
        \]
    \ENDFOR

    \STATE \textbf{Stage 2}: Validator Training
    \FOR{each sample $(u_i,i_i, y_i)$ in ($S, I, Y$) }
        % \STATE \textbf{Step 1:} $G$ generates updated RecCoT $\hat{y}_G = G(x)$.
        \STATE fine-tuning $V$ with objective :
        \[
        % \mathcal{L}_V = \text{Dist}(\hat{y}_G, y)
            % \min_ [y_i(\log V(G(si))+(1-y_i)(\log (1-V(G(s{i}))]
            \max_V \log(\ P_V(y_i|u_i,i_i,G(u_i)\ )
        \]
        % where $\text{Dist}(\cdot)$ is a distance metric (e.g., MSE or cross-entropy).
        % \STATE \textbf{Step 3:} Update $V$ to minimize $\mathcal{L}_V$ by descending gradient:
        \STATE update $V's$ parameters : 
        \[
        V \leftarrow V + \beta \nabla \log(\ P_V(y_i|u_i,i_i,G(u_i)\ )
        \]
    \ENDFOR

    % \STATE \textbf{Check Convergence:} If average feedback score $r > \tau$, terminate training.
\ENDFOR
\STATE \textbf{Output:} Optimized models $G$ and $V$.

\end{algorithmic}

\end{algorithm}

To enhance $G$'s RecCoT reasoning capability, we need to optimize $G$'s uncertainty in reasoning. The specific process is shown in \textbf{Algorithm \ref{alg1}}. Given user historical interaction  $U$, target item $I$, we first adjust $G$'s reasoning parameters, such as temperature $t$ and $top\_p$, and perform $N$ inferences to generate $N$ different RecCoTs $Cot_{ij}$. 
% Formatted as:
% \begin{equation}
% c_{ij} = G(s_i, I_i, t, top\_p)  
% \end{equation}
% where $t$ and $top\_p$ represent the inference parameters temperature and top p, respectively, and $c_{ij}$ denotes the inferred recommendation CoT.

\begin{figure*}[t]
\includegraphics[width=0.98\textwidth]{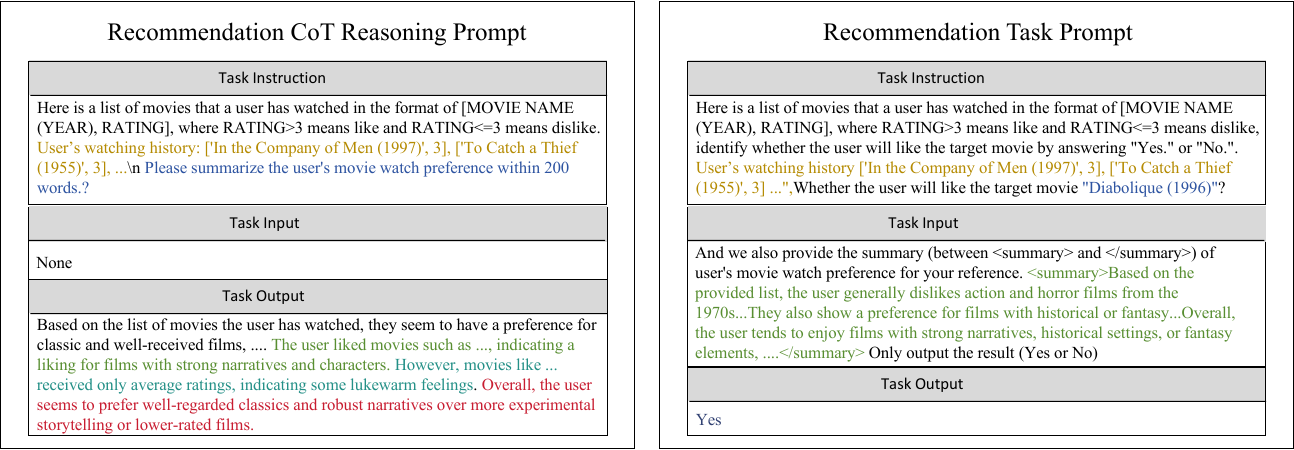}
\caption{Example prompts for TrackRec, including recommendation CoT reasoning prompt denoted as $prompt^{CoT}$ and LLM recommendation task prompt denoted as $prompt^{Rec}$, used for CoT preference alignment and Rec-tuning respectively.}
\label{pprompt}
\end{figure*}

Subsequently, we use $V$ to provide feedback $r_{ij}$ on the $N$-Sampling RecCoTs generated by $G$. We design recommendation task prompt $prompt^{Rec}$ as shown in Figure \ref{pprompt} , for $V$ to perform recommendation tasks. Each RecCoT is incorporated into $V$'s  $prompt^{Rec}$ to perform the recommendation task, obtaining the probabilities of $V$'s output $a_{ij}$ for 'Yes' and 'No', which are then normalized. 

% Formatted as:
% \begin{equation}
%   p^{T}_{ij} = P(a_{ij} = Yes),\ \   p^{F}_{ij} = P(a_{ij} = No), 
% \end{equation}

% \begin{equation}
%     p^{T}_{ij} = \frac{p^{T}_{ij}}{p^{T}_{ij} + p^{F}_{ij}},  p^{F}_{ij} = \frac{p^{F}_{ij}}{p^{T}_{ij} + p^{F}_{ij}}\\  
% \end{equation}
% where $a_{ij}$ denotes the V's output.

\textbf{Feedback Reward}: The feedback reward is calculated by comparing the generated answers with the actual user behavior. This can be formalized as:
\begin{equation}
    r_{ij} = \begin{cases} 
                        p^{T}_{ij} & \text{if } y_i = Yes \\
                        p^{F}_{ij} & \text{if } y_i = No 
                    \end{cases}\\ 
\end{equation}
where $y_i$ denotes the user's actual behavior, such as whether they click or not, $p^{T}_{ij}, p^{F}_{ij}$ represent the probabilities of $V$ outputting $Yes$ and $No$, respectively.

\subsubsection{\textbf{Preference Alignment Based on Sampling Feedback}}
The absence of reference RecCoTs prevents direct supervised fine-tuning of $G$. Furthermore, for multiple sampling, each user interaction history includes one positive sample and multiple negative samples, making it naturally suitable for Softmax-DPO (S-DPO). Therefore, we use S-DPO, a direct preference optimization approach using feedback from the validator $V$. 
% Under the paradigm of reinforcement learning, we formalize the problem of G generating recommendation CoTs into a Markov Decision Process $<S, A, P, r>$, where $S$ represents the state space, $A$ represents the action space, $P$ represents the policy model, and $r$ represents the reward function. In our designed reinforcement learning based on sampling feedback, we can express the above process as:
% \begin{itemize}
%     \item State: All historical interaction behaviors of users and products along with their corresponding Target Items, i.e., the set $S=\{(s_i,I_i)\}_{i=1}^n$, where $n$ represents the total number of $S$.
%     \item Action: The recommendation CoT generator $G$ generates possible recommendation CoTs $c_i$ for each $(s_i, I_i)$, i.e., the set $A_i=\{c_i\}_{i=1\text{'}}^N$, where $N$ represents all possible $c_i$.
%     \item Policy: We initialize the Policy model $\pi_\theta$ using $G$, mapping historical interaction behaviors of users and items to their corresponding recommendation CoTs, i.e., $\pi_\theta=G$.
%     \item Reward: We use the validator $V$ to provide feedback on the recommendation CoT, generating the task answer $a_i$. By comparing $a_i$ with the true label $y_i$, we compute a scalar reward $r_i$. As detailed in Section 3.2.1.
% \end{itemize}
We construct positive RecCoT $Cot_p$ and negative RecCoT set $C^\varepsilon$ for user historical interaction $u_i$ using the feedback reward $r_i$, as follows:
\begin{equation}
    % (c_{chosen}, c_{rejected}) = (c_w, c_l) where r_w > r_l
    % R =\{(c_w,c_l)\mid c_w,c_l\in c_i,r_w>r_l\}
   % c_p = c_i \ where\  r_i = \max\ ( R )\ \\
   \begin{aligned}&C \leftarrow \{{Cot_{i}}\} , R \leftarrow \{r_{i}\}\\&Cot_{p}=Cot_{i}\  where\  r_{i}=\max (R)\\&C^{\varepsilon}=C-Cot_{p}\end{aligned}
\end{equation}
Therefore, the mathematical formula for the optimization objective of S-DPO for $G$ is as follows:
\begin{equation}
    \max_G \log (\ \frac{P_G (\boldsymbol{Cot_{p}|u_{i}})^{N-1}}{ \prod_{Cot_d\in{C^\varepsilon}}P_G(\boldsymbol{Cot_{d}|u_{i}})}\ )
\end{equation}
where $N$ denotes the number of samples.
Therefore, the loss function for optimizing $G$ can be expressed as follows:

\begin{equation}
    \begin{aligned}&\mathcal{L}_{\mathrm{S-DPO}}(\pi_\theta;{G})=-\log \sigma \\& \left(-\log \sum_{Cot_d\in{C^\varepsilon}}\exp\left(\beta\log\frac{\pi_\theta(Cot_d\mid u_i)}{{G}(Cot_d\mid u_i)}-\beta\log\frac{\pi_\theta(Cot_{p}\mid u_i)}{{G}(Cot_{p}\mid u_i)}\right)\right)\end{aligned}
\end{equation}

where $Cot_p$ and $Cot_d$ respectively represent the positive RecCoT and the negative RecCoTs, $\pi_\theta$ denotes the Policy model. Improve the reasoning ability of $G$'s RecCoT by maximizing the distance between the positive RecCoT and multiple negative RecCoTs.

\subsection{Iterative Alternating Feedback Learning}
To further enhance the recommendation CoT generation capability of the LLM and the performance of the recommendation task, we design an iterative alternating learning for $G$ and $V$, as detailed in \textbf{Algorithm \ref{alg2}}. In each iteration of alternating learning, for each user historical interaction sample $u_i$, $G$ generates $N$ recommendation CoTs through multiple sampling. These recommendation CoTs are then evaluated by $V$ to provide feedback, resulting in feedback rewards. Positive sample $Cot_{p}$ and negative sample set $C^\varepsilon$ are constructed based on the feedback rewards, and S-DPO is applied to $G$. After optimization, $G$ performs greedy decoding to generate recommendation CoTs, which are used as feedback to construct an instruction dataset for fine-tuning $V$. 
% The optimization objective is as follows:
% \begin{equation}
%             L_V = \min [y_i(\log V(G(si))+(1-y_i)(\log (1-V(G(s{i}))]
% \end{equation}
% \textbf{In the new iteration of alternating learning, $G$ is derived from the generator after the previous iteration of reinforcement; to avoid overfitting of the validator $V$, $V$ is derived from the base LLM.} 
\textbf{In the new iteration of alternating learning, both $G$ and $V$ are inherited from the preference-aligned generator and validator of the previous iteration.} 
% To prevent resource waste and overfitting due to iterative learning, we refer to the experience of SPIN \cite{chen2024selfplay} and set the number of iterations T to 5.

% \vspace{-10pt} 
\subsection{Rec-tuning}
To enhance $V$'s validation and recommendation capabilities to align with the recommendation task, we design Rec-tuning for $V$. In each iteration of alternating feedback learning, we use $G$, which has been optimized in that iteration, to perform greedy decoding to generate RecCoTs. These CoTs are then assembled with the recommendation task prompt $prompt^{Rec}$ and label to form a recommendation instruction dataset. The example \textit{Recommendation Task Prompt} is shown in Figure \ref{pprompt}.
% \vspace{-15pt} 

We utilize this instruction dataset to fine-tune the validator $V$, with the training objective formulated as follows:
\begin{equation}
    \max_V \log(\ P_V(y_i|u_i,i_i,G(u_i)\ )
\end{equation}
Therefore, the loss function for Rec-tuning can be expressed as follows:
\begin{equation}  
    L_V =  y_i(\log V(u_i,i_i,G(u_i))+(1-y_i)(\log (1-V(u_i,i_i,G(u_{i}))
\end{equation}
where $y_i$ represents the user behavior label, $u_i$ denotes the user's historical interaction, $i_i$ denotes the target item.

\subsection{Distillation}
% The inference quality of LLMs in zero-shot settings is easily affected and constrained by model size. Therefore, we perform zero-shot inference by ChatGPT-4o to generate RecCoTs, constructing a distillation dataset to fine-tune the smaller model, such as Qwen-7B, to enhance its reasoning capability.

% The reasoning quality of LLMs in zero-shot setting is easily affected by model size, and LLMs with larger parameter scales generally possess stronger reasoning capabilities. However, the reasoning efficiency of large-scale LLMs is intolerable for recommendation systems. Therefore, we design a recommendation Chain-of-Thought (RecCoT) distillation module. We select 5\% of the users from the dataset and use a large-scale LLM for zero-shot reasoning to generate RecCoTs, thereby constructing a distillation dataset to fine-tune smaller models, enhancing the reasoning capabilities of smaller LLMs.

The reasoning quality of LLMs in zero-shot settings is highly susceptible to model scale, with larger-parameter models typically exhibiting stronger reasoning capabilities. However, the inference efficiency of large-scale LLMs is intolerable for recommendation systems. To address this, we design a Recommendation Chain-of-Thought (RecCoT) distillation module: We construct a large-scale user sequence prompt dataset incorporating recommendation data (prompt template $prompt^{CoT}$ shown in Figure 4). These user sequence prompts are fed to a powerful teacher model like GPT-4o, and the generated chain-of-thought are combined with the original prompts to create a dataset for fine-tuning smaller student models (e.g., Qwen-7B). Remarkably, we found that using only about 5\% of this data can significantly enhance the reasoning capabilities of smaller LLMs.

\subsection{Preference  Utilization}
For LLM-enhanced recommendation, integrating aligned user preference into the backbone recommendation model is crucial. In this section, we elaborate on how our method utilizes user preferences within the backbone recommendation model. We directly employ a knowledge encoder like Bert \cite{34-bert} to encode user preferences into dense representation as enhanced user features. These enhanced vectors are treated as additional features in the backbone recommendation model. Specifically, we use them as extra feature fields in the recommendation model, enabling explicit interaction with other features. Further, we develop a \textit{connector} $\mathcal{M}$, implemented as a MLP. This connector facilitates the transition of these representations from the semantic space to the recommendation space, simultaneously enabling dimensionality reduction. This process can be formalized as follows:
\begin{equation}
e^{p} = Agg(Encoder(G(u_i))
\label{embedding equ}
\end{equation}
\begin{equation}
    \hat{y} = \mathcal{R}(x,\mathcal{M}(e^p))
\end{equation}
Here, \( x \) represents ID features in traditional recommendation model. $\mathcal{R}$ denotes the backbone recommendation model, and $\mathcal{M}$ denote the connector for user preference. During training, $\mathcal{M}$ is jointly optimized with the backbone model $\mathcal{R}$ via the binary cross-entropy (BCE) loss.
% \subsection{Application to Online Recommendation Systems}
% \begin{figure}
% \centering
% \includegraphics[width=0.48\textwidth]{texts/figure/crop_online_2_.pdf}
% \caption{The pipeline of Online System. We deploy TrackRec on the Ad online platform of Kuaishou}
% \label{fig4}
% \vspace{-3mm}

% \end{figure}
% Our main goal is to apply LLM in large-scale recommendation systems to generate chain-of-thought CoT) suitable for existing recommendation models, thereby providing information gain for the recommendation system. As shown in Figure \ref{fig4}, in the offline training phase, we generate recommendation CoT through multiple sampling based on users' interaction histories with items, and obtain feedback through a validator. Through iterative alternating feedback learning (S-DPO, Rec-tuning), we develop a user preference generator aligned with recommendation tasks. We then use this fine-tuned user preference generator to generate recommendation CoT, embedding the CoT as additional dense features of users and items and storing them in the database. In the online service phase, similar to all other features, we retrieve the user and candidate item features from the database to perform inference in the online advertising recommendation ranking model.

\section{Experiments}
\begin{table}[h]
  \caption{Statistics of datasets used in this paper.}
  \label{tab:statistics}
  \begin{tabular}{l|ll lll llll}
    \toprule
    Dataset & \#Train & \#Valid & \#Test & \#User & \#Item \\
    \midrule
    Amazon-Book & 727468& 25747 & 25747 & 22967 & 34154\\
    MovieLens & 33891 & 10401 & 7331 & 839 & 3256\\
  \bottomrule
  % \hline
\end{tabular}
\end{table}
\begin{table*}
  \caption{Effectiveness Analysis of TrackRec Across Different Backbone CTR Models on MovieLens and Amazon-books Datasets. Qwen-7B as Rec denotes using Qwen-7B for recommendation tasks.}
  \label{tab:public result}
  % \begin{tabular}{|c|c|c|c|c|c|c|c|c|c|c|c|c|}
  \begin{tabular}{cccccccccccccc}
    \toprule
    % \hline
    \multirow{3}{*}{\textbf{Models}}  &\multicolumn{6}{c}{\textbf{Amazon-Book}} &\multicolumn{6}{c}{\textbf{MovieLens}} \\ 
    \cline{2-13}
    & \multicolumn{2}{c}{\textbf{AUC}}& \multicolumn{2}{c}{\textbf{ACC}} &\multicolumn{2}{c}{\textbf{LogLoss}}& \multicolumn{2}{c}{\textbf{AUC}}& \multicolumn{2}{c}{\textbf{ACC}}& \multicolumn{2}{c}{\textbf{LogLoss}} \\
    \cline{2-13}
    & \multicolumn{1}{c}{\textbf{Base}} &\multicolumn{1}{c}{\textbf{Our}}& \multicolumn{1}{c}{\textbf{Base}} &\multicolumn{1}{c}{\textbf{Our}}& \multicolumn{1}{c}{\textbf{Base}} &\multicolumn{1}{c}{\textbf{Our}}& \multicolumn{1}{c}{\textbf{Base}} &\multicolumn{1}{c}{\textbf{Our}}& \multicolumn{1}{c}{\textbf{Base}} &\multicolumn{1}{c}{\textbf{Our}}& \multicolumn{1}{c}{\textbf{Base}} &\multicolumn{1}{c}{\textbf{Our}}\\
    \midrule
    % MLP & 0.x& \textbf{0.x}& 0.x& \textbf{0.x}& 0.x& \textbf{0.x}& 0.x& \textbf{0.x}& 0.x& \textbf{0.x}& 0.x& \textbf{0.x}\\
    % GRU4Rec & 0.x& \textbf{0.x}& 0.x& \textbf{0.x}& 0.x& \textbf{0.x}& 0.x& \textbf{0.x}& 0.x& \textbf{0.x}& 0.x& \textbf{0.x}\\
     AutoInt &0.8071 & \textbf{0.8163\textsuperscript{*}}& 0.7445 & \textbf{0.7482\textsuperscript{*}}& 0.5158& \textbf{0.5057\textsuperscript{*}}& 0.7553& \textbf{0.7628\textsuperscript{*}}& 0.6877 & \textbf{0.6918\textsuperscript{*}}& 0.5873& \textbf{0.5750\textsuperscript{*}}\\
    DeepFM & 0.8086& \textbf{0.8163\textsuperscript{*}}& 0.7451& \textbf{0.7464\textsuperscript{*}}& 0.5154& \textbf{0.5131\textsuperscript{*}}& 0.7556& \textbf{0.7668\textsuperscript{*}}& 0.6868 & \textbf{0.6974\textsuperscript{*}}& 0.5823 & \textbf{0.5749\textsuperscript{*}}\\
    % xDeepFM & 0.8056& \textbf{0.8177}& 0.7365& \textbf{0.7409}& 0.5264& \textbf{0.5083}& 0.7657& \textbf{0.7668}& 0.7020& \textbf{0.7023}& 0.5860 & \textbf{0.5727}\\
    FiGNN & 0.8066& \textbf{0.8188\textsuperscript{*}}& 0.7437& \textbf{0.7448\textsuperscript{*}}& 0.5168& \textbf{0.5079\textsuperscript{*}}& 0.7575& \textbf{0.7688\textsuperscript{*}}& 0.6910& \textbf{0.6972\textsuperscript{*}}& 0.5801 & \textbf{0.5737\textsuperscript{*}}\\
    FiBiNet & 0.8077& \textbf{0.8189\textsuperscript{*}}& 0.7442& \textbf{0.7511\textsuperscript{*}}& 0.5150& \textbf{0.5046\textsuperscript{*}}& 0.7560 & \textbf{0.7691\textsuperscript{*}}& 0.6892 & \textbf{0.6992\textsuperscript{*}} & 0.5843 & \textbf{0.5834\textsuperscript{*}}\\
    DCN & 0.8085& \textbf{0.8181\textsuperscript{*}}& 0.7453& \textbf{0.7505\textsuperscript{*}}& 0.5149 & \textbf{0.5028\textsuperscript{*}}& 0.7563 & \textbf{0.7696\textsuperscript{*}} & 0.6899 & \textbf{0.6954\textsuperscript{*}}& 0.5851 & \textbf{0.5683\textsuperscript{*}}\\
    DIN & 0.8002& \textbf{0.8125\textsuperscript{*}}& 0.7400& \textbf{0.7428\textsuperscript{*}}& 0.5257& \textbf{0.5159\textsuperscript{*}}& 0.7621 & \textbf{0.7729\textsuperscript{*}}& 0.6879 & \textbf{0.6996\textsuperscript{*}}& 0.5798 & \textbf{0.5709\textsuperscript{*}}\\
    DIEN & 0.8030 & \textbf{0.8142\textsuperscript{*}}& 0.7422& \textbf{0.7453\textsuperscript{*}}& 0.5199& \textbf{0.5109\textsuperscript{*}}& 0.7608 & \textbf{0.7710\textsuperscript{*}}& 0.6896 & \textbf{0.6980\textsuperscript{*}}& 0.5825 & \textbf{0.5743\textsuperscript{*}}\\
    
    % MIND & 0.x& \textbf{0.x}& 0.x& \textbf{0.x}& 0.x& \textbf{0.x}& 0.x& \textbf{0.x}& 0.x& \textbf{0.x}& 0.x& \textbf{0.x}\\
    %MIMN & 0.x& 0.x& 0.x& 0.x& 0.x& 0.x& 0.x& 0.x& 0.x& 0.x& 0.x& 0.x\\
    % \hline
     % Qwen-7B & 0.4838& \textbf{}&  0.5571& \textbf{}& 0.7503& \textbf{}& 0.5080& \textbf{0.5607}& 0.4860& \textbf{0.5120}&- & \textbf{-}\\
     \hdashline
    Qwen-7B as Rec & 0.8318 & \textbf{0.8368\textsuperscript{*}}& 0.7563& \textbf{0.7607\textsuperscript{*}}& \textbf{-} & \textbf{-}& 0.7762 & \textbf{0.7887\textsuperscript{*}}& 0.7011 & \textbf{0.7143\textsuperscript{*}}& - & \textbf{-}\\
    % \hline
    
  \bottomrule
\multicolumn{13}{l}{* denotes statistically significant improvement (measured by t-test with p-value<0.001) over baselines. }
\end{tabular}
\end{table*}
% In this section, we present our experiments in detail, including datasets, experimental setup, model comparison, and corresponding analyses. The proposed method is compared with several state-of-the-art works on two public datasets and one industrial dataset. Furthermore, we analyze the effect of each component in our model and prove their usefulness and necessity.
In this section, we conduct extensive experiments on both public and industrial datasets to answer the following questions:
\begin{itemize}[leftmargin=*]
\item {{\bfseries RQ1:}} What improvements can TrackRec bring to backbone models and LLM for recommendation, such as CTR prediction?

\item {{\bfseries RQ2:}}  How does TrackRec perform compared with other LLM-based
baseline methods?

\item {{\bfseries RQ3:}}  Does TrackRec gain performance improvement when deployed online?

\item {{\bfseries RQ4:}} How does preference alignment of TrackRec contribute to performance improvement?
% \end{itemize}

\item {{\bfseries RQ5:}} What improvements can iterative alternating feedback learning bring to recommendation?

\item {{\bfseries RQ6:}}  How does distillation impact recommendation performance?
\end{itemize}

\subsection{Experimental Setup}

\subsubsection {Datasets.}
We conduct experiments on two public datasets and one industrial dataset: {{\bfseries Amazon Dataset\footnote{\url{https://nijianmo.github.io/amazon/index.html}}}} 
% is composed of product reviews and meta-data from Amazon. This is a book recommendation dataset processed from BookCrossing \cite{ziegler2005improving}. The BookCrossing dataset has user ratings (1-10) and textual descriptions of books, such as the information of 'Book-Author' and 'Book-Title'. For each user, we randomly select an item interacted by this user as the prediction target, and sample 10 interacted items as historical interactions. Subsequently, we partition constructed fine-tuning samples into training, validation, and testing sets with the same ratio of 8:1:1. Additionally,
% we binarize the ratings according to a threshold of 5.
is the "Book" category of the Amazon Review Dataset. For this dataset, we regard reviews as one type of interaction behaviors, and sort the reviews from each user by time. {{\bfseries MovieLens Dataset\footnote{\url{https://movielens.org/}}}} 
% contains 1 million ratings provided by 6040 users for 3706 movies. We follow the approach used in previous studies \cite{w25, zhang2023reformulating} by considering 10 interacted items per user as historical interactions. We process the original dataset by sampling the most recent 10,000 interactions and split them into training, validation, and testing sets with a ratio of 8:1:1. To construct a fine-tuning sample, 10 interactions prior to the target item are retained as historical interactions.
is a widely recognized movie recommendation benchmark dataset, provided
by GroupLens research. The dataset comprise user ratings for movies and includes textual information for users and items, such as movie titles. {{\bfseries Industrial Dataset}} is collected from a large advertising platform with hundreds of millions of users. Samples are constructed through sampling from impression logs. 
%The training set is composed of samples from the past several days, and the test set is from the following day, encompassing data from over 400 million users, 10 of millions of advertisements, and tens of billions of behavioral records.

For public dataset processing, we entirely follow the setup of CoLLM \cite{zhang2023collm}, including label processing, data selection, and partitioning methods. Notably, we randomly sample a small portion of data from the training set to initialize the Generator $G$ through distillation, in order to enhance the reasoning ability of the small LLM. The statistics of the processed datasets are presented in Table \ref{tab:statistics}.

% \subsection{Experimental Setup}
\subsubsection {Backbone Models.}
 We focus on critical recommendation tasks, including CTR and CVR prediction. Due to the model-agnostic nature of TrackRec, we select several classic CTR and CVR models and LLM-based prediction models to validate its effectiveness. \textbf{AutoInt} ~\cite{56-autoint} adopts a self-attentive neural network with residual connections to model the interactions explicitly.
\textbf{DeepFM} ~\cite{27-deepfm} use factorization machine to capture low-order and high-order feature interactions.  
% \textbf{xDeepFM} ~\cite{4_xdeepfm} propose a novel Compressed Interaction Network, which aims to generate feature interactions in an explicit fashion and at the vector-wise level.  
\textbf{FiGNN} ~\cite{fignn} design a novel model Feature Interaction Graph Neural Networks
to take advantage of the strong representative power of graphs.
\textbf{FiBiNet} ~\cite{fibinet} is an abbreviation for Feature Importance and Bilinear feature Interaction NETwork is proposed to dynamically learn the feature importance and fine-grained feature interactions.
\textbf{DCN} ~\cite{55-dcn} incorporates cross-network architecture to learn the bounded-degree feature interactions. 
\textbf{DIN} ~\cite{23-din} utilizes attention to model user interests dynamically with respect to a certain item. 
\textbf{DIEN} ~\cite{24-DIEN} introduces an interest evolving mechanism to capture the dynamic evolution of user interests over time. 
\textbf{Qwen-7B} ~\cite{qwen2} can understand users' historical behavior and perform recommendation tasks through zero-shot learning and instruction tuning.
% \begin{itemize}
% \item {{\bfseries GRU4Rec}} ~\cite{22-grurec} models sequential user behaviors based on the Recurrent Neural Network.
% \item {{\bfseries DIN}} ~\cite{9_din} utilizes attention to model user interests dynamically with respect to a certain item. 
% \end{itemize}

\subsubsection {Competitors.}
We compare our work with methods in the same domain that leverage LLMs to enhance recommendations. We divide competitors into two types of models with different settings:
\begin{itemize}[leftmargin=*]
    \item \textbf{LLM as Rec.} \textbf{TALLRec} \cite{38-tallrec} , \textbf{RLPF} \cite{rl1} finetunes LLMs and enhances the recommendation capabilities of LLMs in few-shot scenarios. \textbf{LLM-REC} \cite{44-44} incorporates four distinct prompting strategies for personalized text-based recommendations. \textbf{CoLLM} \cite{zhang2023collm} and \textbf{BinLLM} \cite{binllm} maps user and item representations obtained from collaborative model to LLMs token space.

    \item \textbf{LLM improve Rec.} 
% \item {{\bfseries LLM augmented recommendation model with text input only.}}
\textbf{KAR} \cite{14_kar}, \textbf{RLPF} \cite{rl1} acquires the knowledge about users and items from LLMs as augment for recommendation models. 
\end{itemize}

% , LLM directly outputs predicted recommendation results.

% \subsubsection {LLM Models.}
% We utilize two widely-used LLMs to generate CoT reasoning. Specifically, we employ GPT-3.5 ~\cite{12_gpt4,36-gpt3.5} and ChatGLM-6B ~\cite{57-glm} for 

\subsubsection {Parameter Configuration.}
In the multiple sampling stage, we set the number of samples \( N = 10 \), temperature \(t = 1\) and \(top\_p = 0.9\). Then, we use Qwen2.5-7B\cite{qwen2} as the base model for both the generator \( G \) and the validator \( V \). We use ChatGPT-4o\cite{12_gpt4} for distillation. We set the number of iterations \( T = 3 \). We use the AdamW optimizer with a learning rate of \( 1 \times e^{-5} \) to optimize \( G \) and \( V \). For preference utilization, we use BERT as the Encoder. Other parameters, such as batch size and learning rate, are determined through grid search to achieve the best results. For fair comparisons, the parameters of the backbone model and the baselines are also
tuned to achieve their optimal performance.

\subsubsection {Evaluation Metrics.}
We use the following metrics to evaluate the performance of our proposed model. AUC, ACC and LogLoss (binary crossentropy loss) are almost the default offline evaluation metrics in industry since offline AUC directly reflects the online performance. A higher AUC and ACC value or a lower Logloss value, even by a small margin can be viewed as a significant improvement in CTR and CVR prediction performance.

\subsection{Offline Experimental Results (RQ1, RQ2)}

\subsubsection {Improvement over different backbone Models (RQ1). }
We implement our proposed TrackRec on several representative CTR and CVR prediction models as well as LLM-based prediction models. Specifically, we incorporate the Recommendation CoT generated by G in TrackRec into different prediction models to observe its impact on these various models. Specifically, we use BERT to get RecCoT embedding from LLM inference, and use it as additional information input into the recommendation model. The results are shown in Table \ref{tab:public result}, from which we can draw the following observations: (i) the application of TrackRec significantly improve the performance of various baseline models, validating that the iterative alternating feedback learning framework can continuously enhance the generation quality of RecCoT, thereby improving the performance of recommendation models. (ii) We have observed that LLMs perform better as recommendation models compared to traditional models. We believe one reason for this is that user behavior in public datasets is relatively sparse, which is more suitable for leveraging the reasoning capabilities of LLMs. (iii) As a model-agnostic framework, TrackRec can be applied to a wide range of baseline models, including traditional recommendation models and LLMs. By incorporating our framework, the selected representative models show significant improvements in AUC and ACC on two public datasets, demonstrating the generalizability of TrackRec.
%(iii) Furthermore, we utilize data with limited interactions as the test set to verify the effectiveness of cold starts. Experiments demonstrate that the CRR achieves significant improvements on long-tail data, indicating that the extensive reasoning knowledge of LLM can effectively compensate for the deficiencies of ID-based models in handling long-tail data.

\subsubsection {Superiority: Improvement over Baselines (RQ2).}
\begin{table*}[h]
% \vspace{-3mm}
  \caption{Performance Comparison with LLM-based Recommenders (LLM as Rec and LLM improve Rec) on MovieLens and Amazon-books Datasets. The best results are highlighted in bold, while the second-best results are indicated with underlining. "Rel.Impr." is the relative improvement rate of TrackRec against each baseline.}
  \label{tab:compare baseline}
  % \begin{tabular}{cc|cc|c|c|c|c|c|}
  \begin{tabular}{ccccccccccc}
    \toprule
    % \hline
    \multirow{2}{*}{Setting} & \multicolumn{2}{c}{\multirow{2}{*}{Method}} & \multicolumn{4}{c}{Amazon-Book} &\multicolumn{4}{c}{MovieLens}\\ 
    \cline{4-11}
    &&& AUC & Rel.Impr.& ACC & Rel.Impr. & AUC & Rel.Impr. & ACC & Rel.Impr. \\
    \midrule
     \multirow{6}{*}{LLM as Rec} & \multirow{2}{*}{zero-shot} & TALLRec &0.4857 & 72.28\% &0.5740& 32.52\% & 0.5612 & 40.53\% & 0.5389 & 32.54\%  \\
     && LLMRec & 0.4820 & 73.61\% & 0.4856 & 56.65\% & 0.5320 & 48,25\%& 0.5268 & 35.59\% \\
     % \vspace{0.1mm}
    % \cline{2-11}
     % \vspace{0.1mm}
    % \vspace{1mm}
    &\multirow{2}{*}{sft-based} %&TALLRec &0.7375 & - & 0.7097 & - \\
     & CoLLM & 0.8245 & 1.49\% & 0.7096 & 7.20\% & 0.7296 & 8.10\% & 0.6622 & 7.86\% \\
     && BinLLM & 0.8264 & 1.29\% & 0.7028 & 8.23\% & 0.7425 & 6.22\%& 0.6250 & 14.28\%\\
     
    &  \multirow{1}{*}{RL-based} 
    % & DMPO & - & - & - &- \\
    
    & RLPF & \underline{0.8271} & 1.17\% & \underline{0.7548} & 0.78\% & \underline{0.7855} & 0.41\% & \underline{0.7134} & 0.13\%\\
    % & RLPF & - & - & - & - \\
    % \cline{2-11}
    &\multirow{1}{*}{\textbf{Ours}} & TrackRec & \textbf{0.8368\textsuperscript{*}} & - & \textbf{0.7607\textsuperscript{*}} & - & \textbf{0.7887\textsuperscript{*}} & - & \textbf{0.7143\textsuperscript{*}} & - \\
     
     % \multirow{2}{*}{RL} &TALLRec &0.7375 & -&0.7097 &- \\
     % && CoLLM & 0.8245 & - & 0.7190 & -
     \midrule
    
     \multirow{4}{*}{\shortstack{LLM improve Rec}}&\multirow{1}{*}{zero-shot}&KAR & \underline{0.8048} & 0.96\%& 0.7379 & 0.66\% & 0.7649 & 1.04\% & 0.6932 & 0.92\% \\

    &\multirow{1}{*}{RL-based} & RLPF & 0.8046 & 0.98\% & \underline{0.7418} & 0.14\% & \underline{0.7695} & 0.44\% & \underline{0.6986} & 0.15\%  \\
    &\multirow{1}{*}{\textbf{Ours}} & TrackRec & \textbf{0.8125\textsuperscript{*}} & - & \textbf{0.7428\textsuperscript{*}} & - & \textbf{0.7729\textsuperscript{*}} & - & \textbf{0.6996\textsuperscript{*}} & -\\
    
% \hline

  \bottomrule
\multicolumn{11}{l}{* denotes statistically significant improvement (measured by t-test with p-value<0.001) over baselines. }
\end{tabular}
% \vspace{-3mm}
\end{table*}
% \subsubsection {Improvement over Baselines.}
Next, we compare TrackRec with recent LLMs used for recommendation as baseline, which include two types of methods: LLMs directly used as recommenders (LLM as Rec) and LLMs used to enhance the capabilities of recommenders (LLM improve Rec). In the comparison for LLM improve Rec, we consistently use DIN as the prediction model. The results are shown in Table ~\ref{tab:compare baseline}, which leads to the following observations: (i) For LLM as Rec, fine-tuning LLMs, including SFT and reinforcement learning, can significantly enhance the recommendation performance of LLMs. (ii) For LLM improve Rec, the CoT of LLM reasoning can significantly enhance the performance of traditional recommendation models.  (iii) Our model significantly outperforms other LLM-based baselines. Significant improvements in AUC and ACC are achieved.

\subsection{Online Experimental Results (RQ3)}
\subsubsection{Experimental Setup}
% To validate the effectiveness of TrackRec in real-world scenarios, we conduct online A/B test on Kuaishou's advertising platform.  In online A/B test, the traffic of the whole app is split into ten buckets uniformly.  10\% of traffic is assigned to the online baseline while anther 10\% is assigned to TrackRec.  As revealed in Table \ref{tab:industrial result}, Kuaishou serves over 320 million users daily, and the results collected from 10\% of traffic for several weeks are convincing. For extracting open-world item knowledge, we perform LLM inference across all items. In contrast, extracting user preference knowledge, which involves processing billions of data points, results in high inference costs. Therefore, we tailor our inference strategies to the activity levels of users on the Kuaishou's advertising platform. For active users, we conduct full inference using the LLM. For less active users, since the item knowledge has already been fully inferred, we approximate the inference results of the item knowledge in the user's historical behavior list as the inference results for their preferences.
\begin{table}[h]
% \vspace{}
  \caption{Performance of TrackRec on the industrial dataset, with comparisons conducted on both the long-tail and all aspects.}
  \label{tab:industrial result}
  \begin{tabular}{ccccc}
    % \toprule
    \hline
    Method &Setting &AUC&Revenue&CVR\\ 
    %\cline{2-7}
    %& $AUC$ & $Revenue$& $CVR$& $AUC$ & $Revenue$& $CVR$\\
    \midrule
    \multirow{2}{*}{Base} & long-tail & 0.8172& -& -\\
    \cline{2-5}& all & 0.8335 & -& -\\
    \hline
    \multirow{2}{*}{TrackRec} & long-tail & \textbf{0.8232\textsuperscript{*}}& \textbf{+ 4.2\%}& \textbf{+ 1.7\%}\\
    \cline{2-5}& all & \textbf{0.8363\textsuperscript{*}}& \textbf{+ 2.3\%}& \textbf{+ 1.6\%}

\\ \hline
  %\bottomrule
\multicolumn{5}{l}{* denotes statistically significant improvement (p-value<0.001).}
\end{tabular}
% \vspace{-3mm}
\end{table}

To validate the effectiveness of TrackRec in real-world scenarios, we conduct extensive offline and online experiments on a large advertising platform which serves hundreds of millions of users, millions of advertisements and billions of user interaction records daily. In large-scale recommendation systems, the complexity of large models cannot meet the requirements for online latency. It is impractical to use large models directly as recommendation models. Instead, LLMs are more often used as information reasoners to assist downstream recommendation models in making predictions. Therefore, we train the RecCoT generator $G$ and RecCoT validator $V$ based on the behavioral log information collected by the advertising system. For users, due to the enormous data scale at the billion-level, the inference cost is high, so we only perform LLM inference on active users. The inference results from $G$ will be introduced as side information into the online advertising model for estimation. 

In the online A/B testing, the entire advertising platform's traffic is evenly divided into multiple buckets. More than 10\% of the traffic is allocated to the online base model, and another more than 10\% is allocated to TrackRec.
\subsubsection{Experiment Results.}  As shown in the table \ref{tab:industrial result}, In the experimental group, recommendation CoTs are generated using TrackRec and use as dense features for the ranking model. The control group uses the original ranking model as the baseline. During a 14-day online A/B test, our method show an increase in revenue by \textbf{2.3\%} and an improvement in conversion rate by \textbf{1.6\%} compared to the baseline, resulting in significant business benefits. Furthermore, the world knowledge and reasoning capabilities of the LLM effectively compensate for the recommendation model's shortcomings in long-tail estimation (i.e., cold start). We validate the effectiveness of handling cold starts using data with limited interactions. The experiments demonstrate that TrackRec achieves a revenue increase of \textbf{4.2\%} and a conversion rate improvement of \textbf{1.7\%} on long-tail data, indicating that the broad reasoning capabilities of LLMs effectively address the shortcomings of ID-based models in handling long-tail data.
% \vspace{-3mm}

% \begin{table}[h] 
%     \caption{Comparison of inference efficiency under different QPS (Queries Per Second) in online service.}
%     \label{efficiency}
    
%     \begin{tabular}{ccccc}
%     \hline
%     \diagbox[width=2.2cm]{Model}{QPS}& 100 & 200 & 400 & 1000 \\
%     \midrule
%     base&20.1ms&20.3ms&20.8ms&21.5ms\\
%     TrackRec&20.2ms&20.2ms&21.6ms&21.7ms\\
%     \bottomrule
%     \end{tabular}
%     % }
%     \vspace{-3mm}
% \end{table}
\begin{table}[t] 
\caption{Comparison of efficiency of LLMs and inference efficiency under different QPS (Queries Per Second) in online service.}
    \label{Efficiency}
    % \vspace{-3mm}
    % \resizebox{\textwidth}{!}{
    \centering 
    \begin{subtable}{0.27\textwidth}
    \caption{Efficiency of LLMs.}
%    \caption{Effect of hypergraphs on performance:  \\
%    removing one hypergraph per experiment}
    \label{efficiencya}
    \centering 
    \resizebox{\textwidth}{!}{
    \begin{tabular}{ccc}
        \toprule
          \multirow{2}{*}{Model} & Train Time  & Inference Time \\
          & Per Samples & Per Sample \\
        \midrule
        Qwen-7B & - & 1303ms \\
        % SFT w/ Origin ID & 515ms & 1038ms \\
        TrackRec & 524ms & 613ms \\
        % \midrule
        % Downstream & w/o DR & x &x \\
        % Model & HI-Rec & x & x \\
        \bottomrule
    \end{tabular}
    }
    \vspace{2mm}
    \end{subtable}
    % \hfill
    \begin{subtable}{0.35\textwidth}
    \caption{Efficiency of online model.}
    \label{efficiencyb}
    \centering 
    \resizebox{\textwidth}{!}{
    \begin{tabular}{ccccc}
    \toprule
    \diagbox[width=2.2cm]{Model}{QPS}& 100&200&400&1000\\
    \midrule
    base&20ms&20.1ms&20.8ms&21.5ms\\
    TrackRec&\textbf{20.0ms}&\textbf{20.4ms}&\textbf{21.1ms}&\textbf{21.6ms} \\
    \bottomrule
    \end{tabular}
    }
    \end{subtable}
    % }
    %\vspace{-3mm}
\end{table}

\subsubsection{Efficiency of TrackRec.} 
% For TrackRec, the extraction of RecCoT is performed in an offline manner. In online experiments, the generated RecCoT is directly integrated as auxiliary information into the advertising model, resulting in negligible inference latency. As shown in Table \ref{efficiency}, during online experiments, the inference time for online services show almost no increase across different QPS levels, fully meeting the latency requirements for recommendation services in production.
To evaluate the training and inference efficiency of TrackRec, we conduct comparative experiments using an A800 80G GPU server. During offline LLM inference (as shown in Table \ref{efficiencya}, the Iterative Alternating Feedback Learning mechanism effectively aligns with user preferences, reducing redundant token generation and lowering inference complexity, thereby demonstrating significant efficiency advantages for TrackRec. In online experiments, we maintain a fixed number of serving machines while progressively increasing request volume to observe response times under different QPS levels. As shown in Table \ref{efficiencyb}, in actual deployment on the advertising platform, the inference time for online services show almost no increase across different QPS levels, fully meeting the latency requirements for recommendation services in production. 

% Overall, TrackRec exhibits remarkable inference efficiency advantages that meet the requirements of industrial-grade recommendation systems.

\subsection{Ablation Study}
% \begin{table}[htbp]
% % \vspace{-3mm}

% \centering
% \caption{Ablation study of different Components of our methods. Rec-tuning as the base setting,   $\mathbf{K_{1}}$, $\mathbf{K_{2}}$, $\mathbf{K_{3}}$ represent Distillation, CoT preference alignment and Iterative alternating feedback learning , respectively. }
% \label{ablation study}
% % \vspace{-3mm}
% \begin{tabular}{c|ccc|cc|cc}
% \toprule
% \multirow{2}{*}{Validators} & \multicolumn{3}{c|}{Components} & \multicolumn{2}{c}{Amazon\-Book} & \multicolumn{2}{c}{MovieLens}\\ 
%  & $\mathbf{K_{1}}$ & $\mathbf{K_{2}}$ & $\mathbf{K_{3}}$ & AUC & ACC  & AUC & ACC\\ 
% \hline
% % \multirow{4}{*}{Qwen-7B}
% % & &  &  & 0.7772 & 0.6977        \\ 
% % & \checkmark &  &   & 0.7816 &  0.7074      \\ 
% % &  & \checkmark &   & - &  -     \\ 
% % & \checkmark & \checkmark &   & 0.7825 & 0.7101         \\ 
% % &  &\checkmark  & \checkmark  & - & -         \\ 
% % & \checkmark & \checkmark & \checkmark  & \textbf{0.7836} & 0.7022       \\
% % \hline
% \multirow{4}{*}{DIN}
% & &  &  & 0.7602 & 0.6934 & - &-       \\ 
% & \checkmark &  &   &  &  -  & - & -    \\ 
% & \checkmark & \checkmark &   & - & -  & - & -       \\ 
% & \checkmark & \checkmark & \checkmark  & - &- &- &-     \\

%   \bottomrule                      
% \end{tabular}

% \end{table}

We conduct ablation study on the MovieLens  and  Amazon-Book dataset to evaluate the preference alignment, iterative alternating learning , distillation and Rec-tuning proposed in TrackRec. 
% Previous experiments confirmed the necessity of Rec-tuning for LLM as Rec. For LLM as Rec, our ablation experiments are conducted with the Rec-tuning configuration. For LLM improve Rec, we select DIN as the validation model to compare the performance of different modules.

\begin{figure*}
\includegraphics[width=0.8\textwidth]{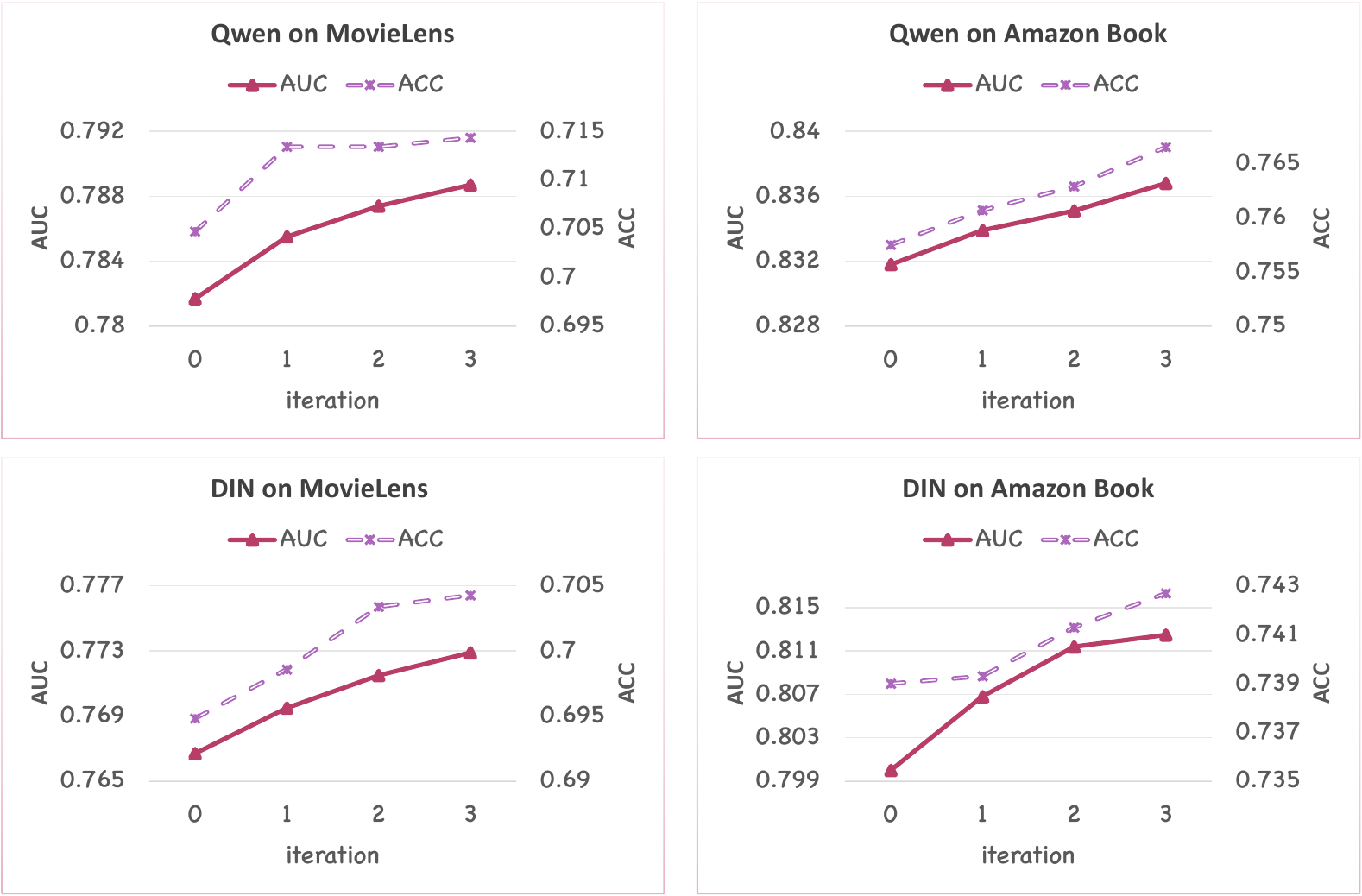}
\caption{Performance of the iterative alternating learning mechanism on different datasets.}
\label{iteration}
\end{figure*}

\subsubsection{Effect of preference alignment (RQ4).} 
\begin{table}[htbp]
\centering
\caption{Impact of preference alignment module $M_{c}$. w/o and w/ refer to without and with $M_{c}$, respectively.}
\label{tab:preference}
\resizebox{\columnwidth}{!}{
\begin{tabular}{ccccccc}
\toprule
\multirow{2}{*}{Datasets} & \multirow{2}{*}{Validators} & \multicolumn{2}{c}{AUC} & \multicolumn{2}{c}{ACC}  \\
\cmidrule(lr){3-4} \cmidrule(lr){5-6}
         & & w/o $M_{c}$ & w/ $M_{c}$ & w/o $M_{c}$ & w/ $M_{c}$    \\
\midrule
\multirow{2}{*}{Amazen Book} & DIN  & 0.8002   &  \textbf{0.8068\textsuperscript{*}}  &  0.7400  &  \textbf{0.7412\textsuperscript{*}}   \\
&  Qwen-7B  & 0.8318   &  \textbf{0.8339\textsuperscript{*}}  &  0.7563  &  \textbf{0.7588\textsuperscript{*}}   \\

\midrule

\multirow{2}{*}{MovieLens} & DIN  & 0.7621   &  \textbf{0.7695\textsuperscript{*}}  &  0.6879  &  \textbf{0.6986\textsuperscript{*}}   \\
&  Qwen-7B  & 0.7762   &  \textbf{0.7855\textsuperscript{*}}  &  0.7011  &  \textbf{0.7134\textsuperscript{*}}   \\
\bottomrule
\multicolumn{6}{l}{* denotes statistically significant improvement (p-value<0.001).}
\end{tabular}
}
\end{table}
In this section, we explore the impact of preference alignment module on experimental results. For our experiments on MovieLens and Amonzen Book, we select Qwen-7B as the Validator in LLM as Rec and DIN as the Validator in LLM improve Rec. Specifically, we compared the performance of models with and without preference alignment in a single-iteration setting on two public datasets. 
% Specifically, we observe the effect of each component on recommendation performance by using each component independently and incrementally adding them one by one. Since iterative alternating feedback learning can not function alone, we utilized the S-DPO module to observe its influence on recommendation results. 

% As shown in Table \ref{ablation study}, each component independently enhances recommendation performance metrics, and progressively adding each component leads to continued improvement in recommendation effectiveness. Specifically, the Distillation module increases AUC by xx\%, highlighting its role in improving the baseline capacity of LLM-generated recommendation CoT. The S-DPO further raises AUC by XX\%, greatly enhancing the LLM's original capability for inferring user preferences. Iterative alternating feedback learning yields an additional xx\% AUC improvement, further boosting the LLM’s recommendation ability and achieving the best performance. These results demonstrate the significant impact of each component in improving recommendation model performance. The strong recommendation performance across different validators further proves the transferability and generalizability of our TrackRec framework.

As shown in Table \ref{tab:preference}, whether the LLM is used directly as a recommender or with a traditional recommendation model, the recommendation performance is improved to varying degrees. Preference alignment lead to an increase in AUC for Qwen-7B by 0.25\% and 1.19\% on two public datasets, and an increase for DIN by 0.82\% and 0.97\%. These results indicate that integrating a preference optimization module enables the model to more effectively focus on accurately and reasonably inferring user preferences, thereby enhancing the performance of recommendation models.

% \begin{table}[htbp]
% % \vspace{-3mm}
% \centering
% \caption{Performance comparison of various reinforcement learning method for one iteration on the MovieLens Dataset.}
% \vspace{-3mm}
% \label{tab:different rl}
% \begin{tabular}{cccc}
% \toprule
% \multirow{2}{*}{Validators} & \multirow{2}{*}{\begin{tabular}[c]{@{}c@{}}Method\end{tabular}} & \multicolumn{2}{c}{Metrics} \\
% \cmidrule(lr){3-4}
% & & AUC & ACC  \\
% \midrule
% \multirow{5}{*}{Qwen-7B } & None & 0.7548 & 0.686 \\
% & MCPG \cite{mcpg} & 0.7710 & 0.670 \\
% & KTO \cite{kto} & 0.762 & 0.696 \\
% & DPO \cite{dpo} & 0.7686 & 0.707 \\
% & S-DPO & \textbf{0.7736} & \textbf{0.708} \\

% \bottomrule

% \end{tabular}
% \vspace{-3mm}
% \end{table}
% \subsubsection{Effect of iterative alternating learning (RQ5).}
% \begin{figure}
% \centering
% \includegraphics[width=0.48\textwidth]{texts/figure/iteration-已裁剪.pdf}
% \caption{Discuss the issues related to introducing chain-of-thought (CoT) into recommendation systems, including Challenge 1: Unreliable of CoT reasoning, Challenge 2: Weak reasoning ability of small LLM.}
% \label{fig2}
%    % \vspace{-3mm}

% \end{figure}

In this section, we further explore the impact of iterative alternating learning on the performance of recommendation systems. We conduct 3 iterations on the MovieLens and Amonzen Book dataset using Qwen-7B and DIN as the validator, respectively. Iteration 0 indicates the performance when the LLM makes recommendations directly under the initial RecCoT generated by $G$ and incorporates this as auxiliary information into the recommendation model.

As shown in Figure \ref{iteration}, the recommendation performance of the model gradually improves with each iteration, reaching its best performance in the third iteration. On the MovieLen dataset, the LLM used directly as a recommender achieves an improvement by 0.90\%. The DIN recommender achieves an improvement of 0.81\%. On the Amazon Book dataset, Qwen-7B and DIN achieve improvements by 0.60\% and 1.56\%, respectively. These results indicate that iterative alternate feedback learning continuously enhances the ability of large language models to generate effective recommendation chain-of-thought (CoT) reasoning, thereby improving their overall recommendation performance.

\subsubsection{Effect of distillation (RQ6).}
\begin{table}[h]
\centering
\caption{Impact of distillation module $M_{d}$ on recommendation performance. w/o and w/ refer to without and with $M_{d}$, respectively.}
\label{tab:diss}
\resizebox{\columnwidth}{!}{
\begin{tabular}{ccccccc}
\toprule
\multirow{2}{*}{Datasets} &\multirow{2}{*}{Backbones} & \multicolumn{2}{c}{AUC} & \multicolumn{2}{c}{ACC}  \\
\cmidrule(lr){3-4} \cmidrule(lr){5-6} 
          & & w/o $M_{d}$ & w/ $M_{d}$ & w/o $M_{d}$ & w/ $M_{d}$    \\
\midrule
\multirow{2}{*}{Amazon Book} & DIN  & 0.8002   &  \textbf{0.8019\textsuperscript{*}}  &  0.7400  &  \textbf{0.7411\textsuperscript{*}}   \\
& Qwen-7B  & 0.8318   &  \textbf{0.8333\textsuperscript{*}}  &  0.7563  &  \textbf{ 0.7588\textsuperscript{*} }   \\

\bottomrule

\multirow{2}{*}{MovieLens} & DIN  & 0.7621   &  \textbf{0.7667 \textsuperscript{*}}  &  0.6879  &  \textbf{0.6948 \textsuperscript{*}}   \\
& Qwen-7B  & 0.7772   &  \textbf{0.7817 \textsuperscript{*}}  &  0.7011  &  \textbf{ 0.8047 \textsuperscript{*}}   \\
\bottomrule
\multicolumn{6}{l}{* denotes statistically significant improvement (p-value<0.001).}
\end{tabular}
}
\end{table}
In this section, we explore the impact of the distillation module in TrackRec on recommendation performance. We compared the performance of using large language models (LLMs) directly as recommenders and using the traditional recommendation model DIN as a recommender, both with and without the distillation module, across two public datasets. As shown in Table \ref{tab:diss}, on the MovieLens dataset, using Qwen-7B directly as a recommender achieve an improvement by 0.56\%. When using DIN as a recommender, DIN achieve an improvement by 0.33\%. On the Amazon Book dataset, Qwen-7B and DIN achieved improvements of 0.84\% and 0.21\%, respectively. These results indicate that the distillation module can significantly enhance the inference capabilities of small-scale LLMs, thereby improving their recommendation performance.

\section{Conclusions}
Our work presents TrackRec, an iterative alternating feedback learning framework, to continuously improve the user preference inference and recommendation capabilities of LLMs, integrating them into large-scale recommendation systems. Specifically, we decompose the recommendation task using LLMs into two components: the RecCoT Generator $(G)$, responsible for generating user preferences, and the RecCoT Validator (V), which performs recommendation tasks based on user preferences. Both components are derived from small-scale models. First, we design a model distillation module that leverages recommendation prompts to obtain RecCoT generated by a larger LLM and uses it to fine-tune \( G \), initializing the generator to enhance the reasoning capabilities of small-scale models. Additionally, through Rec-tuning and preference alignment, we iteratively improve \( G \)'s user preference reasoning capability and \( V \)'s validation capability, bridging the gap between LLMs and recommendation tasks.

Our work demonstrates superior performance compared to state-of-the-art methods. Moreover, we share practical insights into implementing this framework in a large advertising platform. This method has been deployed on a large advertising platform serving hundreds of millions of users, resulting in a revenue increase of 2.3\% and a conversion rate improvement of 1.6\%.

\bibliographystyle{ACM-Reference-Format}
\balance
\bibliography{main}

\appendix

\end{document}